\let\va=\varphi
\let\la=\lambda
\let\al=\alpha
\let\ra=\rightarrow
\let\disp=\displaystyle
\let\ti=\times

\let\si=\sigma

\let\De=\Delta
\let\pa=\partial

\let\ov=\overrightarrow
\let\ovl=\overline
\def\que#1#2{\displaystyle\frac{#1}{#2}}

\let\si=\sigma

\documentclass[pra]{revtex4}

\usepackage{amsfonts,amssymb,bm,graphicx}

\begin{document}

\title{Preliminary analysis of the possibility of making use of part of the
energy flow of zero-point radiation}

\author{R. Alvargonz\'alez and L. S. Soto}

\affiliation{Facultad de F\'{\i}sica,  Universidad Complutense,
28040 Madrid, Spain}

\date{\today}

\begin{abstract}
The energy flow of zero-point radiation is very great, but difficult to put
to use. However, the observations made by Sparnaay in 1958 and by Lamoureux in
1997 reveal the possibility of making use of a very small fraction of that
immense amount. 
This possibility is big enough for such a minute fraction to have significant
importance, but such a possibility requires miniaturisation to a degree which
may be unattainable. It is worth trying to achieve it, since it would open the
way to interstellar travel.
\end{abstract}

\maketitle

\section{Introduction}

In 1958, Sparnaay discovered that two uncharged conductor plates, situated
in a near perfect vacuum and at a temperature near absolute zero, attract each
other with a force whose intensity per surface unit is inversely proportional
to the fourth power of the distance $d$ which exist between them. For
$d=5\ti10^{-5}$ cm, Sparnaay was able to measure a force of $0.196$
g$\cdot $cm$\cdot $s$^{-2}$, and deduced the formula: $f=\que{k_s}{d^4}$, in which
$k_s$, Sparnaay's constant, is equal to $1.225\ti10^{-18}$ erg$\cdot $cm.

In [1], we proposed the hypothesis that the force measured by Sparnaay is
caused by the fact that the presence of the two uncharged conductor plates
prevents the photons of zero-point radiation which fall at an angle greater
than $\va$ and have a wavelength greater than $d/\cdot os\va$, from being
reflected from the space between the two plates, since they will have been
already reflected from outside that space. We can see from Fig. 1 that the
`do not fit''into the space between the plates.

\begin{figure}[h]
\centering
\resizebox{0.70\columnwidth}{!}{\includegraphics{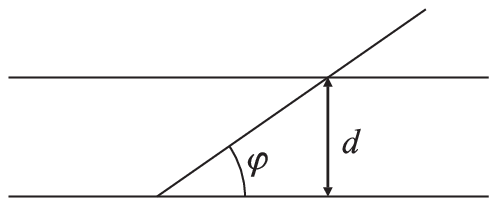}}
\caption{Fig. 1}
\end{figure}

In a perfect vacuum and at the absolute temperature $T=0$ K$^{\rm o}$, the
phenomenon observed by Sparnaay can only be due to a radiation inherent to
space, the existence of which had been suggested by Nernst in 1916. Such a
radiation would have to possess a relativistically invariant spectrum, which
implies that the relative abundance of its photons must be inversely
proportional to the cubes of
their wavelengths. In other words, the number of photons of
wavelength $\la$ which fall on a given area, within a given lapse of time,
must be inversely proportional to $\la^3$.

A radiation whose spectrum is such that the relative abundance of its photons
is inversely proportional to the cubes of their wavelengths, implies a
distribution of energy which is inversely proportional to their fourth power.
In 1969, Timothy H. Boyer showed that the function of spectral density of
zero-point radiation is:
$$f_\va(\la)=\que1{2\pi^2}\cdot \que1{(\la_*)^3},$$
where $\la_*$ is the number expressing the measurement of wavelength $\la$
[1]. This function means that the energy of the flow of photons of wavelength
$\la$ is:
$$E_\va(\la)=\que1{2\pi^2}\cdot \que{hc}\la\cdot \que1{(\la_*)^3}.$$
For $\la\ra0$, $E_\va(\la)\ra\infty$. There must therefore exist a minimum
wavelength, $q_\la$, to which there corresponds the maximum frequency
$\que1{q_\tau}$ such that $q_\la/q_\tau=c$.

Starting from the hypothesis proposed in [1], we reached in that paper the
equation (6):
$$f=\que{2\De W}c=\que{\rho\pi}{\al x}\cdot \que1{d^4}\que{m_el_e}{t_e^2}\qquad
\hbox{per}\quad l_e^2,$$
where $\rho$ is the coefficient of reflection,
$m_e$ is the mass of the electron, and $l_e$ and $t_e$ are respectively
the units of length and time in the $(e,m_e,c)$ system, in which the basic
magnitudes are electron charge, electron mass and the speed of light. Finally,
$d$ is the distance between the plates, expressed in $l_e$, and $x$ the amount
of $q_\tau$, which elapse between the arrival of one photon of wavelength
$q_\la$, and the arrival of the next one (equation (3) in Ref. [1]).
From this equation, we arrived, in [2], at:
$$f=\que\rho x\cdot 0.407513\ {\rm g}\cdot {\rm cm}\cdot {\rm s}^{-2}.\eqno{(1)}$$
(last equation on p. 3 of [2]).

By equating $F$ with Sparnaay's observations, we obtain:
$\que\rho x\cdot 0.407513=0.196$. Since $1>\rho>0$, the only values possible for
$x$ are $x=1$ and $x=2$. For $x=1$, $\rho=0.4810$; for $x=2$, $\rho=0.962$,
which is improbable for wavelengths near to $5\ti10^{-5}$ cm.

If we incorporate $x=1$ into (1), the number 0.407513 must correspond to:
$\que{K\pi}\al\cdot \que1{(5\ti10^{-5})^4}=0.407513$, whence
$$K=\que{0.407513(5\ti10^{-5})^4\al}\pi=5.91612\ti10^{-21},$$
and the last equation on p. 3 of [2] can be written as:
$$f=5.91612\ti10^{21}\que{\rho\pi}\al\cdot \que1{d^4}\ {\rm g}\cdot {\rm cm}\cdot {\rm
s}^{-2},\eqno{(2)}$$
in which $d$ is the distance between the plates in cm. When we introduce the
numbers $\pi$ and $\al$,
$$f=2.54670\ti10^{-18}\que\rho{d^4}\ {\rm g}\cdot {\rm cm}\cdot {\rm s}^{-2};$$
for $f=\que{1.225\ti10^{-18}}{d^4}$ g$\cdot $cm$\cdot $s$^{-2}$ we obtain
$\rho=0.4809$.

\section{Theoretical scheme for making use of part of the energy flow of
zero-point radiation}

Sparnaay had to make his measurements with uncharged plates, because if
they had been charged, the electrostatic repulsion between them would have
affected the issue. Keeping this in mind, we can imagine the theoretical sheme
shown in Fig. 2, which would provide a basic elementary model, composed of two
conductor plates $(ab)$ and $(cd)$, moving along the guide-rails
$\ovl{A_1A_2}$; $\ovl{B_1B_2}$ and $\ovl{A_2A_3}$; $\ovl{B_2B_3}$,
respectively.

\begin{figure}[h]
\centering
\resizebox{0.70\columnwidth}{!}{\includegraphics{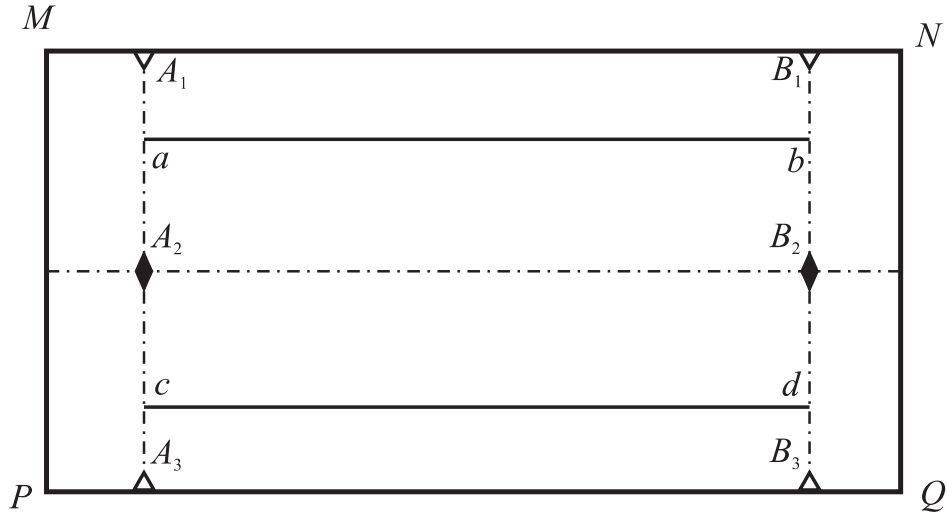}}
\caption{Fig. 2. Scheme for making use of a very small fraction of the energy
flow of zero-point radiation.}
\end{figure}

\vskip 3pt
\noindent{\tabcolsep=0pt\begin{tabular}{ccc}
M & & N\\[-3pt]
& \fbox{\rule[-1pt]{0mm}{3mm}
\rule{3mm}{0mm}}&\\[-3pt]
P & & Q\end{tabular}}\ \qquad Supports (must be placed in a vacuum)

\vskip 3pt
\noindent
$|\ \ |$\\[-3pt]
$|\ \ |$\qquad\ Guide-rails along which the conducting plates
 ab and cd move\\[-3pt]
$|\ \ |$

\vskip 3pt
\noindent$\!\!\!\!\left.\begin{array}{l}
\nabla A_1B_1\\
\Delta A_3B_3\end{array}\right\}$\qquad Sensors inducing the conductor plates
to discharge $\diamondsuit  A_2B_2$\qquad Sensors inducing the conductor plates to
charge \vspace*{2pc}

When the plates arrive near to the guide-rail $\ovl{A_2B_2}$, they will be
very close together, $d_1=10^{-5}$ cm, and sensors will connect each of them
with a condenser during the time needed for electrostatic repulsion to move
than apart, overcoming Sparnaay's forces and defeating the inertia inherent to
their mass and velocity. Arriving respectively one of them close to
$\ovl{A_1B_1}$, and the other close to $\ovl{A_3B_3}$, different sensors will
connect them to a line of discharge, at which Sparnaay's forces will come into
play, causing them to approach each other. It is at this part of the cycle,
when the plates move towards each other, where we could make use of a small
fraction of the energy flow of zero-point radiation.

The scheme may include a system for recovering the electric charge at
$\ovl{A_1B_1}$, and at $\ovl{A_3B_3}$.

\section{Kinematics and dynamics of Sparnaay's attraction,
 between two parallel uncharged plates, situated in a vacuum at 0 K}

Assuming that the plates are square, and a system of reference in which
their centres move along the axis $\ovl{OZ}$, in such a way that when the
centre of one plate is at a distance of $0.5\si$ above $O$, the centre of the
other plate is at a distance of $-0.5\si$ below $O$, there would for all
intents exist between them a mutual attraction expressable by:
$$f_s=\que{1.225\ti10^{-18}}{\si^4}(A) {\rm g}\cdot {\rm cm}\cdot {\rm s}^{-2},$$
where $A$ is the area of the plates in cm and $\si$ the distance between them
in cm.

Copper plates with sides of 1 cm, and thickness of 0.1 mm, have a mass of
$8.95\ti10^{-2}$~g. A force $f_s$ would give them an acceleration of:
$$a_s=\que{1.225\ti10^{-18}}{\si^4}\que{{\rm g}\cdot {\rm cm}\cdot {\rm
s}^{-2}}{8.95\ti10^{-2}}=\que{1.368715\ti10^{-17}}{\si^4}{\rm cm}\cdot {\rm
s}^{-2}\eqno{(3)}$$

This leads to $d\si=atdt=\que{1.368715\ti10^{-17}}{\si^4}tdt$, whence
$\si^4d\si=1.368715\ti10^{-17}tdt$. By integrating, we obtain:
$$\int^{d1}_{d2}\si^4d\si=1.368715\ti10^{-17}\int^{t_1}_{t_2}tdt,$$
and, finally:
$$t_2^2-t_1^2=\que25\que1{1.368715\ti10^{17}}[(d_2)^5-(d_1)^5]\eqno{(4)}$$

Taking as a starting time $t=0$ and $d_1$, we obtain:
$$t^2=2.922449\ti10^{16}[(d_2)^2-(d_1)^5]s^2.$$
For $\si_1=10^{-5}$ cm, we obtain:
$$t=1.709517\ti10^8[(\si_2)^5-10^{-25}]^{1/2}s.\eqno{(5)}$$

From $\disp\int\si d\si=1.368715\ti10^{-17}\int tdt$;
$\que{\si^5}5=1.368715\ti10^{-17}\que{t^2}2,$ whence
$$t=\left(\que25\ti\que1{1.368715\ti10^{-17}}\right)^{1/2}\si^{5/2}=
1.709517\ti10^8\si^{5/2}.$$
From this we derive:
$$dt=\que52\cdot 1.709517\ti10^8\cdot \si^{3/2}d\si=4.273793\ti10^8\si^{3/2}d\si$$
$$v=\que{d\si}{dt}=\que1{\si^{3/2}\ti4.273793\ti10^8}\eqno{(6)}$$

The energy obtained through the movement between $d_2$ and $d_1$ is given by:
$$E^{d_2}_{d_1}=\int^{d_2}_{d_1}f\si d\si=1.225\ti10^{-18}
\int^{d_2}_{d_1}\que{d\si}{\si^4}{\rm g}\cdot {\rm cm}^2\cdot {\rm s}^{-2}.$$
By integration we obtain:
$$E^{d_2}_{d_1}=\que{1.225\ti10^{-18}}3\left[\que1{(\si_1)^3}-\que1{(\si_2)^3}\right].$$

For $\si_1=10^{-5}$ cm,
$$E^{d_2}_{d_1}=4.083\ti10^{-4}-\que{4.083\ti10^{-19}}{(\si_2)^3}.\eqno{(7)}$$

The energy flow per second is given by:
$$\va=\que{E^{d_2}_{d_1}}t{\rm g}\cdot {\rm cm}^2\cdot {\rm s}^{-3}=
\left[4.083\ti10^{-4}-\que{4.083\ti10^{-19}}{(d_2)^4}\right]\cdot  t\cdot {\rm g}\cdot 
{\rm cm}^2\cdot {\rm s}^{-3},\eqno{(8)}$$
where $t$ is the time, in seconds, which has elapsed during the movement
between $d_1$ and $d_2$.

Table 1 gives an overview of the values of $t,v,a,E^{d_2}_{d_1}$ and $\va$, as
functions of the maximum separation of $d_2$~cm, always assuming that
$d_1=10^{-5}$~cm.

\vskip 6pt
\begin{center} TABLE 1\protect\\[+5pt]
$\begin{array}{|c|c|c|c|c|c|}
\hline
d_2\ {\rm (cm)} & t\ {\rm (s)} & v\ {\rm (cm}\cdot {\rm s}^{-1}) &
a\ {\rm (cm}\cdot {\rm s}^{-2}) & E\ {\rm (g}\cdot {\rm cm}^2{\rm s}^{-2}) &
\va={\rm g}\cdot {\rm cm}^2{\rm s}^{-3}\\
\hline
10^{-3} & 5.406 & 7.40\ti10^{-5} & 1.369\ti10^{-5} & 4.083329\ti10^{-4} &
7.553\ti10^{-5}\\
\hline
8\ti10^{-4} & 3.094 & 1.03\ti10^{-4} & 3.341\ti10^{-5} & 4.083325\ti10^{-4} &
1.320\ti10^{-4}\\
\hline
6\ti10^{-4} & 1.507 & 1.59\ti10^{-4} & 1.056\ti10^{-4} & 4.083314\ti10^{-4} &
2.710\ti10^{-4}\\
\hline
4\ti10^{-4} & 0.547 & 2.92\ti10^{-4} & 5.347\ti10^{-4} & 4.083269\ti10^{-4} &
7.465\ti10^{-4}\\
\hline
2\ti10^{-4} & 9.67\ti10^{-2} & 8.27\ti10^{-4} & 8.554\ti10^{-3} &
4.083229\ti10^{-4} & 4.221\ti10^{-3}\\
\hline
10^{-4} & 1.71\ti10^{-2} & 2.34\ti10^{-3} & 1.369\ti10^{-1} &
4.079250\ti10^{-4} & 2.385\ti10^{-2}\\
\hline
8\ti10^{-5} & 9.79\ti10^{-3} & 3.27\ti10^{-3} & 3.342\ti10^{-1} &
4.075358\ti10^{-4} & 4.163\ti10^{-2}\\
\hline
6\ti10^{-5} & 4.77\ti10^{-3} & 5.03\ti10^{-3} & 1.056 &
4.064429\ti10^{-4} & 8.521\ti10^{-2}\\
\hline
6\ti10^{-5} & 4.77\ti10^{-3} & 5.03\ti10^{-3} & 1.056 &
4.064429\ti10^{-4} & 8.521\ti10^{-2}\\
\hline
4\ti10^{-5} & 1.73\ti10^{-3} & 9.25\ti10^{-3} & 5.346 &
4.019531\ti10^{-4} & 2.323\ti10^{-1}\\
\hline
2\ti10^{-5} & 3.06\ti10^{-4} & 2.62\ti10^{-2} & 8.554\ti10^{-1} &
3.572916\ti10^{-4} & 1.168\\
\hline
\end{array}$\end{center}\vskip 1pc

\section{Kinematics and dynamics of the electrostatic repulsion between the
two parallel plates, ${ab}$ with charge $\va_1$ and
${cd}$ with charge $\va_2$}

When the two parallel plates $ab$ and $cd$ of Fig. 2 are each connected to
condensers, the first plate to a condenser with a potential of $\va$ {\em
statvolts}, and the other to one with a potential of $\va_2$ {\em
statvolts}, the configuration composed of both plates proceeds to act as a
single vacuum condenser made up to flat plates, of plate area $A$~cm$^2$, a
distance of $\si$~cm betweem the plates, and a potential difference of
$(\va_1-\va_2)$ {\em statvolts} $=\va_{12}$ {\em statvolts}. The electrostatic
repulsion between the charge $Q_1$ of plate $ab$ and the charge $Q_2$ of plate
$cd$ drives them to separate with a force of $\que{Q_1Q_2}{\si^2}$. This force
disappears when both plates are discharged, when the distance between them is
$\si_2$. Both $A$ and $\va_1$ and $\va_2$ remain constant throughout the
repulsion, the only magnitude which varies during the repulsion being $\si$.
We will consider the characteristics of a vacuum condenser of area $A$
(cm$^2$), potential difference between the plates of $\va_{12}$ ({\em
statvolts}) and a distance between the plates of $\si$ (cm), wich are:
\begin{itemize}\itemsep=0pt
\item { Capacity} $C$ (cm) $=\que{Q(ues)}{\va_{12}(statvolt)}$\ \ ([4]\
p. 100), which may also be expressed as
$$C\,({\rm cm})=\que{A\,({\rm cm}^2)}{4\pi\si\,({\rm cm})}\qquad\hbox{([4]\
p. 103)}$$\eject
\item {lectric Field}\qquad

$E\,({\rm g}^{1/2}{\rm cm}^{-1/2}{\rm
s}^{-1})=\que{\va_{12}({\rm g}^{1/2}{\rm cm}^{1/2}{\rm s}^{-1})}{\si\,({\rm
cm})}$\qquad ([4]\ p. 103)
\item {Charge}\qquad $Q\,({\rm g}^{1/2}{\rm cm}^{3/2}{\rm s}^{-1})=
C\,({\rm cm})\va_{12}({\rm g}^{1/2}{\rm cm}^{1/2}{\rm s}^{-1})$\qquad([4]\ p.
103)
\item {Stored Energy}\qquad$U\,({\rm g}\,{\rm cm}^2{\rm s}^{-2})=
\que12C\,({\rm cm})\cdot (\va_{12})^2\,({\rm g}\,{\rm cm}\,{\rm s}^{-2}),$
which can also be expressed as
$$U\,({\rm g}\,{\rm cm}^2{\rm s}^{-2})=
\que{A}{8\pi\si}{\rm cm}(\va_{12})^2\,({\rm g}\,{\rm cm}\,{\rm s}^{-2}).$$
\item {Separating Force}\qquad $F\,({\rm g}\,{\rm cm}\,{\rm s}^{-2})=
Q^2({\rm g}\,{\rm cm}^3{\rm s}^{-2})/\si^2({\rm cm}^2)$
\item {Acceleration of Movement}\qquad
$$a({\rm cm\,s}^{-2})=
F\,({\rm g}\,{\rm cm}\,{\rm s}^{-2})/(8.95\ti10^{-2})({\rm g})$$
\end{itemize}

In this last formula, the quotient $8.95\ti10^{-2}$ (g) is the mass of the
plate of area $1\ {\rm cm}^2$ and thickness 0.1 mm.

Assuming these identities, and following the system already used for analyzing
the kinematics and dynamics of the plates subjected to Sparnaay's forces, we
can write:
$$a\,{\rm cm\, s}^{-2}=Q^2\,({\rm g}\,{\rm cm}^3{\rm s}^{-2})\cdot \si^{-2}
{\rm cm}^{-2}/(8.95\ti10^{-2}{\rm g})=$$
$$=\left(\que A{4\pi\si}\right)^2\que{(\va_{12})^2\si^{-2}}{(8.95\ti10^{-2})}
{\rm cm\,s}^{-2}=
\que{A^2(\va_{12})^2}{(8.95\ti10^{-2})16\pi^2}\ti\que1{\si^4}.$$
For $A=1\,{\rm cm}^2$, we obtain:
$$a\,{\rm cm\,s}^{-2}=\que{(\va_{12})^2\si^{-4}}{(8.95\ti10^{-2})\ti16\pi^2}
{\rm cm\,s}^{-2},$$
which is only a function of $\si$ when $\va_{12}$ is constant, as would happen
in the mechanism illustrated in Fig. 2.

Also, $v=\que{d\si}{dt}=at$, and since
$\left\{\que{(\va_{12})^2}{(8.95\ti10^{-2}16\pi^2}\right\}$ is constant, we
can write:
$$\que{d\si}{dt}=\{\ \}\si^{-4}\ti t,\qquad\hbox{whence}\ \ \si^4d\si=\{\ \}tdt.$$
By integrating, we obtain:
$$\int^\si_{\si_1}\si^4d\si=\{\ \}\int^t_{t_1}tdt\ra
\que15\Bigl[\si^5\Bigr]^{\si_2}_{\si_1}=
\que12\{\ \}\Bigl[t^2\Bigr]^t_{t_1}.$$

Considering that $\si=\si_1$ corresponds to $t=0$, we arrive at:
$$t=\left(\que{2(\si)^5}{5\{\ \}}\right)^{1/2}-
\left(\que{2(\si_1)^5}{5\{\ \}}\right)^{1/2}=
\que{(2/5)^{1/2}(8.95\ti10^{-2})^{1/2}\ti4\pi}{\va_{12}}
(\si^{5/2}-\si^{5/2}_1){\rm s}.$$

For $\va_{12}=10^{-4}st_v$, $\si_1=2\ti10^{-5}$
$$t=2.377669\ti10^4[\si^{5/2}-1.788854\ti10^{-12}]{\rm s}\eqno{(9)}$$

For $\si=10^{-3}{\rm cm}$; \ $t=7.519\ti10^{-4}$ s.

Assuming that $t=\que{2.377669}{\va_{12}}(\si^{5/2}-\si_1^{5/2})$, we obtain:
$$dt=\que{2.377669}{\va_{12}}\cdot \que52\si^{3/2}d\si,\qquad\hbox{whence:}$$
$$\que{d\si}{dt}=\que25\cdot \que{\va_{12}}{2.377669}\si^{-3/2}=
\que{\va_{12}}{5.944173}\cdot \si^{-3/2}\eqno{(10)}$$

For $\va_{12}=10^{-4}stv$, $\que{d\si}{dt}=1.68232\ti10^{-5}\si^{-3/2}$, which
for $\si_2=10^{-3}$~cm gives $v=0.5319$ cm\,s$^{-1}$.

The formula to find the energy stored in a vacuum condenser;
$U\,({\rm g}\,{\rm cm}^2{\rm s}^{-2})=\que A{8\pi\si}{\rm cm}\cdot 
(\va_{12})^2\,({\rm g}\,{\rm cm}\,{\rm s}^{-2})$ \
where $A=1$ cm$^2$, is
$U\,({\rm g}\,{\rm cm}^2{\rm s}^{-2})=\que{(\va_{12})^2}{8\pi\si}
\,({\rm g}\,{\rm cm}^2{\rm s}^{-2}),$
where $\va_{12}$ is expressed in {\em statvolts} and $\si$ in cm\,s.
Since $\va_{12}$ is constant, this energy decreases as the distance between
the plates increases. So that for $\si=10^{-3}$~cm, it is 50 times smaller
than for a starting value of $\si_1=2\ti10^{-5}$~cm. This decrease implies a
progressive restoration of the charge to the condensers which are connected
respectively to plates $ab$ and $cd$ of Fig. 2, so that the energy used to
move the plates back to a distance of $\si$ is given by:
$$U=\que{(\va_{12})^2}{8\pi\si}\,({\rm g}\,{\rm cm}^2{\rm s}^{-2}).$$
For $\va=10^{-4}stv$ and $\si=10^{-3}$ cm; $U=3.978873\ti10^{-7}
\,({\rm g}\,{\rm cm}^2{\rm s}^{-2}).$

The advantage gained by increasing the maximum distance between the plates
implies, however, the disadvantage of an increase in the value of $t$, which
translates into a decrease in the intensity of the average energy flow
$\phi\,({\rm g}\,{\rm cm}^2{\rm s}^{-3})=U\,({\rm g}\,{\rm cm}^2{\rm s}^{-2})/
ts$.

Table 2 shows values of $t$, $v$, $U$ and $\phi$, for distances between the
plates of between $\si=10^{-3}$ cm and $\si_1=2\ti10^{-5}$ cm. If we compare
them with Table 1, we can see that for $\va_{12}=10^{-4}stv$, the values of
$E$ in Table 1 are higher than those for $U$ in Table 2, but the values for
$\phi$ in Table 2 are lower than those for $\va$ in Table 1. This is because
the duration of the journeys $\ov{s_1s}$, driven by electrostatic impulse, is
much less than that of the journeys $\ov{ss_1}$, driven by zero-point
radiation. Despite this, there is always a positive balance of energy. For
$\va_{12}=10^{-5}stv$, both the energy flows and the energy surpluses in
Table~1 are always higher than those in Table~2.

\begin{center}  TABLE 2\protect\\[+3pt]
1. Values of $t, v, a, U$ and $\phi$ as functions of the maximum
distance between plates, where the minimum distance is $\si_1=2\ti10^{-5}$cm,
$A=1$cm, and $\va_{12}=10^{-4}$  stv \protect\\[+1pc]
$\begin{array}{|c|c|c|c|c|c|}
\hline
\si\ {\rm (cm)} & t\ {\rm (s)} & v\ ({\rm cm\,s}^{-1}) & a\ {\rm (cm\,s}^{-2})
& U\ {\rm (g\,cm}^2{\rm s}^{-2}) & \phi\ ({\rm g\, cm}^2{\rm s}^{-3})\\
\hline
10^{-3} & 7.519\ti10^{-4} & 5.320\ti10^{-1} & 7.0755\ti10^2 & 3.9789\ti10^{-7}
& 5.292\ti10^{-4}\\
\hline
8\ti10^{-4} & 4.304\ti10^{-4} & 7.487\ti10^{-1} & 1.7274\ti10^3 &
4.9736\ti10^{-7} & 1.1557\ti10^{-3}\\
\hline
6\ti10^{-4} & 2.097\ti10^{-4} & 1.1447 & 5.4595\ti10^3 &
6.6314\ti10^{-7} & 3.1623\ti10^{-3}\\
\hline
4\ti10^{-4} & 7.609\ti10^{-5} & 2.1029 & 2.7639\ti10^4 &
9.9472\ti10^{-7} & 1.3073\ti10^{-2}\\
\hline
2\ti10^{-4} & 1.345\ti10^{-5} & 5.9479 & 4.4222\ti10^5 &
1.9894\ti10^{-6} & 1.479\ti10^{-1}\\
\hline
10^{-4} & 2.378\ti10^{-6} & 1.6823\ti10 & 7.0755\ti10^6 &
3.9789\ti10^{-6} & 1.673\\
\hline
8\ti10^{-5} & 1.361\ti10^{-6} & 2.3511\ti10 & 1.7274\ti10^7 &
4.9736\ti10^{-6} & 3.654\\
\hline
6\ti10^{-5} & 6.630\ti10^{-7} & 3.6198\ti10 & 5.4595\ti10^7 &
6.6315\ti10^{-6} & 1.0002\ti10\\
\hline
4\ti10^{-5} & 2.406\ti10^{-7} & 6.6499\ti10 & 2.7639\ti10^8 &
9.9472\ti10^{-6} & 4.1343\ti10\\
\hline
2\ti10^{-5} & 4.253\ti10^{-8} & 1.8809\ti10^2 & 4.4222\ti10^9 &
1.9894\ti10^{-5} & 4.677\ti10^2\\
\hline
\end{array}$\protect\\[+3pc]

 2. Values of $ t, v, a, U$ and $\phi$ as functions of the maximum
distance between plates, where the minimum distance is $\si_1=2\ti10^{-5}$cm,
$A=1$cm, and $\va_{12}=10^{-5}$ stv \protect\\[+1pc]
$\begin{array}{|c|c|c|c|c|c|}
\hline
\si\ {\rm (cm)} & t\ {\rm (s)} & v\ ({\rm cm\,s}^{-1}) & a\ {\rm (cm\,s}^{-2})
& U\ {\rm (g\,cm}^2{\rm s}^{-2}) & \phi\ ({\rm g\, cm}^2{\rm s}^{-3})\\
\hline
10^{-3} & 7.519\ti10^{-3} & 5.320\ti10^{-2} & 7.0755 & 3.9789\ti10^{-9}
& 5.292\ti10^{-7}\\
\hline
8\ti10^{-4} & 4.304\ti10^{-3} & 7.487\ti10^{-2} & 1.7274\ti10 &
4.9736\ti10^{-9} & 1.1557\ti10^{-6}\\
\hline
6\ti10^{-4} & 2.097\ti10^{-3} & 1.1447\ti10^{-1} & 5.4595\ti10 &
6.6314\ti10^{-9} & 3.1623\ti10^{-6}\\
\hline
4\ti10^{-4} & 7.609\ti10^{-4} & 2.1029\ti10^{-1} & 2.7639\ti10^2 &
9.9472\ti10^{-9} & 1.3073\ti10^{-5}\\
\hline
2\ti10^{-4} & 1.345\ti10^{-4} & 5.9479\ti10^{-1} & 4.4222\ti10^3 &
1.9894\ti10^{-8} & 1.479\ti10^{-4}\\
\hline
10^{-4} & 2.378\ti10^{-5} & 1.6823 & 7.0755\ti10^4 &
3.9789\ti10^{-8} & 1.673\ti10^{-3}\\
\hline
8\ti10^{-5} & 1.361\ti10^{-5} & 2.3511 & 1.7274\ti10^5 &
4.9736\ti10^{-8} & 3.654\ti10^{-3}\\
\hline
6\ti10^{-5} & 6.630\ti10^{-6} & 3.6198 & 5.4595\ti10^5 &
6.6315\ti10^{-8} & 1.0002\ti10^{-2}\\
\hline
4\ti10^{-5} & 2.406\ti10^{-6} & 6.6499 & 2.7639\ti10^6 &
9.9472\ti10^{-8} & 4.1343\ti10^{-2}\\
\hline
2\ti10^{-5} & 4.253\ti10^{-7} & 1.8809\ti10 & 4.4222\ti10^7 &
1.9894\ti10^{-7} & 4.677\ti10^{-1}\\
\hline
\end{array}$\end{center}\vskip 2pc

\section{Synthesis and conclusions}

Considering Tables 1 and 2 together, we arrive at the following synthesis:
\def\labelenumi{\arabic{section}.\arabic{enumi})}
\begin{enumerate}\itemsep=0pt
\item The periods of time which correspond to electrostatic repulsion
approximate to $1.4\ti10^{-3}t_s$ for $\va_{12}=10^{-4}stv$, and
$1.4\ti10^{-2}t_s$ for $\va_{12}=10^{-5}stv$, where $t_s$ is the time during
which the plates are moving towards each other, through the action of
zero-point radiation. We can therefore consider, as a first approximation fo
the duration of each complete cycle, that the time taken by the movement of
the plates towards and away from each other is:
\begin{itemize}\itemsep=0pt
\item $t_s(1+1.4\ti10^{-3})=1.707517\ti10^8(1.0014)\si^{5/2}{\rm s.}=
1.709908\si^{5/2}{\rm s.}$, for $\va_{12}=10^{-4}stv$, where $\si$ is the
distance between the plates in cm.
\item $t_s(1+1.4\ti10^{-2})=1.731422\ti10^8\si^{5/2}$ s., for
$\va_{12}=10^{-5}stv.$
\end{itemize}
\item The energy produced by the action of zero-point radiation is given by:
$$E_s=1.225\ti10^{-18}\left(\que1{\si_1^3}-\que1{\si^3}\right);$$
which is approximately equal to $4.083\ti10^{-4}
\,{\rm g}\,{\rm cm}^2{\rm s}^{-2}$ for values of $\si$ greater than
$10^{-4}$~cm. It can therefore be considered as constant for possible
mechanisms where the maximum distance is $\si_2>10^{-4}$~cm.
\item The energy consumed during electrostatic repulsion is given by
$U=\Bigl((\va_{12})^2/8\pi\si_2\Bigr)\,{\rm g}\,{\rm cm}^2{\rm s}^{-2}$, where
$\si_2$ is the maximum separation between the plates. In Tables 1 and 2 we
assumed that $\si_2=10^{-3}$~cm, and for that value $U$is equal to $E_s$ for:
$$ \que{(\va_{12})^2}{8\pi\ti10^{-3}}=4.083\ti10^{-4}.\qquad\
\hbox{It follows that:}$$
$$\va_{12}=(8\pi\ti10^{-3}\ti4.083\ti10^{-4})^{1/2}stv= 3.203\ti10^{-3}stv.$$
Because of this it was possible to establish Table 2 for
$\va_{12}=10^{-4}stv$, and for $\va_{12}=10^{-5}stv$, potential differences
which lead to positive balances of energy, very near to~$E_s$.
\item The energy flow for the ``cycle of approximation and repulsion"\ is
given by:
$$\phi\cdot ong\que{E_s-U}{t_s}\,{\rm g}\,{\rm cm}^2{\rm s}^{-3}=
\que{4.083\ti10^{-4}-(\va_{12})/8\pi\si}{1.710\ti10^8\si^{5/2}}.$$
In the mechanism in Fig. 1, $\va_{12}$ is constant, and the only variable is
$\si$.
$$\phi=2.387719\ti10^{-12}\si^{-5/2}-
2.326827\ti10^{-10}\si^{-7/2}(\va_{12})^2$$
$$\que{\pa\phi}{\pa\si}=\left(-\que52\right)\cdot 2.387719\ti10^{-12}\si^{-7/2}\cdot 
2.326827\ti10^{-10}\si^{-9/2}(\va_{12})^2$$

For $\que{\pa\phi}{\pa\si}=0$, \
$\left(-\que52\right)\cdot 2.387719\ti10^{-12}\si=
\left(-\que72\right)\cdot 2.326827\ti10^{10}.$
$$\si=\left(\que75\right)\que{2.326827\ti10^{-10}}{2.387719\ti10^{-12}}=
1.364297\ti10^2(\va_{12})^2$$

For $(\va_{12})=10^{-4}stv,\qquad \si=1.364297\ti10^{-6}$ cm.

For $(\va_{12})=10^{-5}stv,\qquad \si=1.364297\ti10^{-8}$ cm.

{\arraycolsep=1pt$$\begin{array}{ccl}
\que{\pa^2\phi}{\pa\si^2} & = & \que{35}4\cdot 2.387719\ti10^{-12}\si^{-9/2}-
\left(\que{36}4\right)2.326827\ti10^{-10}\si^{-11/2}(\va_{12})^2\\ \\
& = & \que{\si^{-11/2}}4\Bigl(8.857017\ti10^{-11}\si-
8.376575\ti10^{-9}(\va_{12})^2\Bigr)
\end{array}$$}

$\que{\pa^2\va}{\pa\si^2}>0$ \ both for $\si=1.364297\ti10^{-6}$~cm;
$\va_{12}=10^{-4}stv$ \ and for \ $\si=1.364297\ti10^{-8}$~cm;
$\va_{12}=10^{-5}stv$. These values of $\si$ therefore correspond to maximum
values of $\phi$ and, in both cases, to distances $\si_2$ much smaller than
those considered in this preliminary analysis.
\end{enumerate}\vskip 4pt

In his excellent work ``Special Relativity", A. P. French [3] demons\-tra\-ted the
impossibility in practice of making interstellar journeys by the use of
rockets composed of photons. The technical possibility of using a fraction of
the energy flow of zero-point radiation opens the way for us to hope that such
journeys could be made. (A very long-term prospect, given the nature of the
technological problems inherent in harnessing powerful sources of energy based
on accumulations of mechanisms similar to that shown in Fig. 1.)

\section*{REFERENCES}

[1] R. Alvargonz\'alez: {\em arXiv}: physics/0311027.

[2] R. Alvargonz\'alez: {\em arXiv}: physics/0311139.

[3] A. P. French: {\em Special Relativity}.

[4] E. M. Purcell: {\em Berkeley physics course, Volume 2}.
\end{document}